# Gate control of lattice-pseudospin currents in graphene on WS$_2$: Effect of sublattice symmetry breaking and spin-orbit interaction

Kitakorn Jatiyanon [a], Bumned Soodchomshom[a,*]

[a]Department of Physics, Faculty of Science, Kasetsart University, Bangkok 10900, Thailand

**Abstract**

Strong spin-orbit interaction (SOI) in graphene grown on tungsten disulfide (WS$_2$) has been recently observed, leading to energy gap opening by SOI. Energy gap in graphene may also be induced by sublattice symmetry breaking (SSB) where energy level in A-sublattice is not equal to that in B-sublattice. SSB-gap may be produced by growing graphene on hexagonal boron nitride or silicon carbide. In this work, we investigate transport property in a SOI/SSB/SOI gapped graphene junction, focusing the effect of interplay of SOI and SSB. We find that, lattice-pseudospin polarization (L-PSP) can be controlled perfectly from +100% to -100% by gate voltage. This is due to the fact that in graphene grown on WS$_2$, the carriers carry lattice-pseudo spin degree of freedom "up and down". The SSB-gapped graphene exhibits pseudo-ferromagnetism to play the role of lattice-pseudospin filtering barrier. It is also found that the SOI and SSB-gaps in graphene may be measured by characteristic of L-PSP in the junction. The proposed controllable-lattice-pseudospin currents may be applicable for graphene-based pseudospintronics.

**Keywords:** Graphene; spin-orbit interaction; tungsten disulfide; pseudospintronics

## 1. Introduction

Graphene, a two-dimensional single layer of graphite, has become one of promising 2-dimensional materials for future technology [1-4]. Graphene is formed by carbon atoms arranged in a honeycomb lattice consisting of two sublattices A and B in a unit cell. Undoped graphene behaves like a gapless semiconductor [5]. The conduction and valence bands of graphene touch at precisely two kinds of inequivalent corners $k$ and $k'$ of the first Brillouin zone, yielding its carriers to have valley degree of freedom [6]. The carriers possess the 2-dimensional Dirac fermion where the wave functions of electron in A- and B- sublattices and the Fermi velocity $v_F \cong 10^6 m/s$ play the role of lattice-pseudospin states and the effective speed of light, respectively. Control of spin and valley currents is significant for applications of spintronics and valleytronics, respectively. Graphene may be induced into ferromagnetism by mean of proximity effect [7-11]. Control of spin currents is possible in a ferromagnetic graphene junction [12-16]. The valley-polarized currents in graphene junction may be generated when local strain and magnetic vector potential are applied into graphene junction [17-19]. The application of lattice-pseudospin, pseudospintronics, has also been interested [20]. The highest transverse lattice-pseudospin component occurs in gapped graphene system when the energy of electron approaches the bottom of the band dispersion to generate the so-called "pseudo-ferromagnet" [21]. The energy gap opening in graphene at $k$ and $k'$ points is required to study transport affected by lattice-pseudospin [21-25]. In previous studies of pseudo-magnetotransport in monolayer [21-24] and bilayer [25] graphene junctions, although lattice-pseudospin currents were predicted to be polarized, lattice-pseudospin polarization cannot be controlled by external force. In case of graphene with energy gap $E_{gap}$, the carriers possess the massive Dirac fermions with effective mass obeying the relativistic-like relation $m = E_{gap}/2v_F^2$. The breaking of sublattice-symmetry-induced gap may be created by means of substrate-induced-gap methods [26-29]. Firstly reported, graphene grown on silicon carbide (SiC) has been found to generate energy gap of 260 meV [26] while graphene grown on hexagonal boron nitride (h-BN) may yield the energy gap of about 56 meV [27]. Very recently, the large energy gap 0.5 eV produced in graphene grown on SiC(0001) surface has been reported by Nevius et al [29]. Large energy gap is significant for applying graphene as a good semiconductor.

In general, spin-orbit interaction (SOI) intrinsically present in graphene is too weak [30-33]. Spin-orbit interaction may produce quantum spin Hall state at the edge of graphene applicable for topological electronic devices. Also it may yield the so-called spin-orbit-interaction gap in which its property is different from sublattice-breaking-symmetry gap. The various methods to enhance the magnitude of SOI in graphene have been proposed [34-37]. Avsar et al. [38], have reported that an artificial interface between single layer graphene and few-layer tungsten disulfide ($WS_2$) may lead graphene to acquire spin-orbit-interaction up to 17 meV. Very recently, substrate-induced SOI of about 5 meV has been experimentally observed in graphene on $WS_2$ substrate [39]. The success to amplify SOI in graphene by substrate interaction using $WS_2$ [38, 39] may be applicable for several devices such as spin- and valley-current based devices.

In this paper, transport property in the system of graphene on $WS_2$ substrate is investigated where sublattice-symmetry-breaking (SSB) gap is assumed into the barrier. The SSB-gapped graphene may be induced via interaction by substrates such as SiC [29] and h-BN [27]. The SOI-gapped graphene may be induced by $WS_2$ substrate [39]. In this work, the control of lattice-pseudospin currents is focused. Induced directly by the interplay of SOI and SSB gaps, we will show how lattice-pseudospin currents can be controlled completely by gate voltage.

**2. Hamiltonian model**

In this section, we begin with tight-binding Hamiltonian based on ref.[40] including the effect of spin orbit interactions $\Delta_{SOI}$, Rashba spin-orbit interaction $\Delta_R$ and strength of breaking of energy level in A- and B-sublattices $\Delta_{SSB}$ as given by

$$\hat{h} = \sum_{\langle ij \rangle \alpha} t c_{i\alpha}^\dagger c_{j\alpha}^\dagger + i\frac{\Delta_{SOI}}{3\sqrt{3}} \sum_{\langle\langle ij \rangle\rangle \alpha,\beta} v_{ij} \tau_{\alpha\beta}^z c_{i\alpha}^\dagger c_{j\beta}^\dagger + i\frac{2\Delta_R}{3} \sum_{\langle\langle ij \rangle\rangle \alpha,\beta} \lambda_i c_{i\alpha}^\dagger \left(\vec{\tau} \times \hat{d}_{ij}\right)_{\alpha\beta}^z c_{j\beta}^\dagger$$
$$+ \Delta_{SSB} \sum_{i,\alpha} \lambda_i c_{i\alpha}^\dagger c_{i\alpha}^\dagger + U_G \sum_{i,\alpha} c_{i\alpha}^\dagger c_{i\alpha}^\dagger,$$

(1)

where $t \approx -2.7\,meV$ and $U_G$ are graphene's hoping energy and gate-induced potential energy, respectively. $c_{i(j),\alpha(\beta)}^\dagger$ is a creation operator of electron with spin polarization $\alpha(\beta)$ at site $i(j)$. $\langle i,j \rangle$ and $\langle\langle i,j \rangle\rangle$ stand for summation overall the nearest and next-nearest-neighbor hoping sites, respectively. We have $v_{ij}=1(-1)$ if the next-nearest-neighbor atom is counterclockwise (clockwise) with respect to the z-axis and $\lambda_i=1(-1)$ for A-(B-) sublattice. $\hat{d}_{ij}$ represents unit vector connecting atom $i$ and $j$ in the same sublattice. $\vec{\tau}=<\tau^x,\tau^y,\tau^z>$ and $\vec{\sigma}=<\sigma^x,\sigma^y,\sigma^z>$ are Pauli spin matrix vector acting on **realspin** and **lattice-pseudospin** of electron, respectively. Graphene grown on WS$_2$ in ref.[39] has found that $\Delta_{SOI}\approx 5$ meV and $\Delta_R\approx 1$ meV. In the limit of $\Delta_R\ll\Delta_{SOI}$, the effect due to Rashba spin orbit interaction is neglected. Hence, the low-energy Hamiltonian used to describe the quasiparticle in this system may be given by [39, 41]

$$\hat{H}=v_F\eta\hat{p}_x\sigma^x+v_F\hat{p}_y\sigma^y+\eta s\Delta_{SOI}\sigma^z+\Delta_{SSB}\sigma^z+U, \qquad (2)$$

where $\eta=1(-1)$ and $s=1(-1)$ stands for $k(k')$ valley and realspin $\uparrow(\downarrow)$, respectively. $\hat{p}_{x(y)}=-i\hbar\partial_{x(y)}$ is the momentum operator. The Hamiltonian above would act on wave states called "lattice-pseudospin states", as given by

$$\hat{H}\psi_{\eta s}=E\psi_{\eta s}, \qquad (3)$$

where

$$\psi_{\eta s}=\begin{pmatrix}\psi_A\\\psi_B\end{pmatrix}\equiv\begin{pmatrix}\psi_\Uparrow\\\psi_\Downarrow\end{pmatrix}. \qquad (4)$$

The excited energy is denoted by $E$. The wave states of electron in A-(B-) sublattice $\psi_{A(B)}$ may be considered as equivalent to the amplitude of lattice-pseudospin $\Uparrow(\Downarrow)$ in the z-direction or wave states of electron in A-(B-) sublattices [20].

### 3. Scattering process in a SOI/SSB/SOI gapped graphene junction

In our model, a SOI/SSB/SOI gapped graphene junction is depicted in Fig.1(a). The SOI-regions may be provided by growing graphene on WS$_2$ substrate, while the SSB-region with thickness L may be provided by growing graphene on h-BN or SiC substrate. The current in the junction is assumed to flow in the x-direction. Using the data given by ref.[39], in graphene grown on WS$_2$ carrying the electronic properties that $\Delta_{SOI}\neq 0$ and $\Delta_{SSB}=0$. The Hamiltonian describing electronic field in the SOI-regions, region-1 and 3, may be therefore given by

$$\hat{H}_{1,3} = v_F \eta \hat{p}_x \sigma^x + v_F \hat{p}_y \sigma^y + \eta s \Delta_{SOI} \sigma^z. \tag{5}$$

Graphene grown on h-BN [27] or SiC [29] substrate may give rise to $\Delta_{SOI} = 0$ and $\Delta_{SSB} \neq 0$. The Hamiltonian in the SSB-region, region-2, is thus given as

$$\hat{H}_2 = v_F \eta \hat{p}_x \sigma^x + v_F \hat{p}_y \sigma^y + \Delta_{SSB} \sigma^z + U_G. \tag{6}$$

The wave function describing the scattering process in the system may be given by solving the wave equations above. The wave functions in region-1, 2 and 3 are respectively obtained as

$$\psi_1 = [\begin{pmatrix} 1 \\ Ae^{i\eta\theta} \end{pmatrix} e^{ikx} + r \begin{pmatrix} 1 \\ -Ae^{-i\eta\theta} \end{pmatrix} e^{-ikx}]e^{ik_\parallel y},$$

$$\psi_2 = [a\begin{pmatrix} 1 \\ Be^{i\eta\beta} \end{pmatrix} e^{iqx} + b \begin{pmatrix} 1 \\ -Be^{-i\eta\beta} \end{pmatrix} e^{-iqx}]e^{ik_\parallel y},$$

$$\psi_3 = [t\begin{pmatrix} 1 \\ Ae^{i\eta\theta} \end{pmatrix} e^{ikx}]e^{ik_\parallel y},$$

where $A = \dfrac{E - \eta s \Delta_{SOI}}{\eta\sqrt{E^2 - (\Delta_{SOI})^2}}$, $B = \dfrac{E - U_G - \Delta_{SSB}}{\eta\sqrt{(E - U_G)^2 - (\Delta_{SSB})^2}}$

$$k = \frac{\sqrt{(E)^2 - (\Delta_{SOI})^2}}{\hbar v_F} \cos(\theta) \text{ and } q = \frac{\sqrt{(E - U_G)^2 - (\Delta_{SSB})^2}}{\hbar v_F} \cos(\beta).$$

$$\tag{7}$$

The conservation component is defined by the momentum in the y-direction of the form

$$k_\parallel = \frac{\sqrt{(E)^2 - (\Delta_{SOI})^2}}{\hbar v_F} \sin(\theta) = \frac{\sqrt{(E - U_G)^2 - (\Delta_{SSB})^2}}{\hbar v_F} \sin(\beta), \tag{8}$$

which is used to determine the angle of incidence in the barrier region. By matching the wave function in eq. 7 with the boundary condition of $\psi_1(x=0) = \psi_2(x=0)$ and $\psi_2(x=L) = \psi_3(x=L)$, the coefficients $r$, $a$, $b$ and $t$ can be calculated. By doing this we get the spin-valley-dependent transmission coefficient, given below

$$t_{\eta s}(\theta) = \frac{ABe^{-i(q-k)L}(1 + e^{2i\eta\beta})(1 + e^{2i\eta\theta})}{ABe^{i\eta(\beta+\theta)} - (A^2 + B^2)e^{i\eta(\beta+\theta)}(e^{2iqL} - 1) + AB(1 + e^{2i(qL+\eta\beta)} + e^{2i(qL+\eta\theta)})}.$$

$$\tag{9}$$

## 4. Conductance and lattice-pseudospin polarization

In this section, using the Landauer formula [42], the spin-valley-dependent conductance formulae would be given by the integration overall the angle of incidence, given by

$$G_{\eta s} = \frac{1}{8} G_0 \frac{\sqrt{E^2 - \Delta_1^2}}{E} \int_{-\pi/2}^{\pi/2} T_{\eta s}(\theta) \cos(\theta) d\theta , \quad (10)$$

where $T_{\eta s}(\theta) = t_{\eta s}^*(\theta) t_{\eta s}(\theta)$ is transmission. $G_0 = \frac{4e^2}{h} N_0(E)$ is unit conductance with $N_0(E) = \frac{w}{\pi \hbar v_F} |E|$ being the density of state for gapless graphene. $w$ is the width of graphene sheet, and $N(E) = \frac{w}{\pi \hbar v_F} \sqrt{E^2 - \Delta_{SOI}^2}$ is density of state at the transport channel for gapped graphene. Furthermore, the total conductance G can be calculated using the summation of all spin-valley conductances, $G = G_{k\uparrow} + G_{k\downarrow} + G_{k'\uparrow} + G_{k'\downarrow}$.

Recalling the realspin- and valley-polarizations, they may be usually defined respectively as

$$RSP = \frac{G_\uparrow - G_\downarrow}{G} \times 100\% = \frac{(G_{k\uparrow} + G_{k'\uparrow}) - (G_{k\downarrow} + G_{k'\downarrow})}{G} \times 100\%,$$

and

$$VP = \frac{G_k - G_{k'}}{G} \times 100\% = \frac{(G_{k\uparrow} + G_{k\downarrow}) - (G_{k'\uparrow} + G_{k'\downarrow})}{G} \times 100\%,$$

(11)

where $G_\uparrow = G_{k\uparrow} + G_{k'\uparrow}, G_\downarrow = G_{k\downarrow} + G_{k'\downarrow}$ and $G_k = G_{k\uparrow} + G_{k\downarrow}, G_{k'} = G_{k'\downarrow} + G_{k'\downarrow}$. As we have seen the definition of realspin- and valley-polarizations, it is easy to understand their properties. The net-current in the junction is carried by four current groups, $I_{k\uparrow}, I_{k\downarrow}, I_{k'\uparrow}$, and $I_{k'\downarrow}$. In this section we are going to talking about the lattice-pseudospin currents that may yield the lattice-pseudospin polarization (LSP) defined in similar way as *RSP* and *VP*.

In the model, we may consider that carriers of the system are governed by electron in graphene grown on $WS_2$, while the region-2, graphene grown on h-BN or SiC-substrate region, is just a filtering barrier. To consider the carrier's lattice pseudospin state by beginning with plane wave electron with momentum $<\pm k, k_\parallel>$ valley $\eta = k, k'$ and spin $s = \uparrow, \downarrow$, wave function in graphene grown on WS$_2$ may be taken to be of the form

$$\psi_{\eta s, \pm k} \approx \begin{pmatrix} 1 \\ \pm A e^{\pm i\eta\theta} \end{pmatrix} e^{\pm ikx + ik_\parallel y}.$$

The expectation value of lattice-pseudospin may be given by a usual quantum formula

$$\left\langle \vec{s}_{LP\text{-}spin} \right\rangle_{\eta s, \pm k} = \left\langle \psi_{\eta s, \pm k} \left| \hat{\vec{s}}_{LP\text{-}spin} \right| \psi_{\eta s, \pm k} \right\rangle.$$

To get

$$\left\langle \vec{s}_{LP\text{-}spin} \right\rangle_{\eta s, \pm k} = \pm \eta \frac{\hbar}{2} \sqrt{1 - \left(\frac{\Delta_{SOI}}{E}\right)^2} \left[ \cos(\eta\theta)\hat{i} \pm \sin(\eta\theta)\hat{j} \right] + \frac{\eta s \hbar}{2} \frac{\Delta_{SOI}}{E} \hat{k}.$$

(12)

Here, the lattice pseudo spin operator is given as $\hat{\vec{s}}_{LP\text{-}spin} = \hbar \vec{\sigma}/2$ and $\hat{i}, \hat{j}$ and $\hat{k}$ are unit vectors along the x-, y- and z-directions, respectively. We will see that, in the limit of $E/\Delta_{SOI} \to 1$, expectation value of lattice-pseudospin may be obtained as

$$\left\langle \vec{s}_{LP\text{-}spin} \right\rangle_{\eta s, \pm k} \to \eta s \frac{\hbar}{2} \hat{k}, \qquad (13)$$

to get

$$\left\langle \vec{s}_{LP\text{-}spin} \right\rangle_{k\uparrow} = \left\langle \vec{s}_{LP\text{-}spin} \right\rangle_{k'\downarrow} \to +\frac{\hbar}{2} \hat{k} \equiv \left| \Uparrow \right\rangle$$

$$\left\langle \vec{s}_{LP\text{-}spin} \right\rangle_{k\downarrow} = \left\langle \vec{s}_{LP\text{-}spin} \right\rangle_{k'\uparrow} \to -\frac{\hbar}{2} \hat{k} \equiv \left| \Downarrow \right\rangle.$$

(14)

This is very surprise, because in the regime of $E/\Delta_{SOI} \to 1$ electron in $k$-valley with spin $\uparrow$ and electron in $k'$-valley with spin $\downarrow$ exhibits lattice-pseudospin $\Uparrow$, while electron in $k$-valley with spin $\downarrow$ and electron in $k'$-valley with spin $\uparrow$ exhibits lattice-pseudospin $\Downarrow$. By this way, the current in the junction may be separated into $I_\Uparrow = I_{k\uparrow} + I_{k'\downarrow}$ and $I_\Downarrow = I_{k\downarrow} + I_{k'\uparrow}$ where the net current is $I = I_\Uparrow + I_\Downarrow$. This result would allow us to define lattice-pseudospin polarization in the electronic junction in similar way as *RSP* and *VP*, given of the form

$$L\text{-}PSP = \frac{G_\Uparrow - G_\Downarrow}{G} \times 100\% = \frac{(G_{k\uparrow} + G_{k'\downarrow}) - (G_{k\downarrow} + G_{k'\uparrow})}{G} \times 100\%,$$

(15)

where $G_\Uparrow = G_{k\uparrow} + G_{k'\downarrow}$ and $G_\Downarrow = G_{k\downarrow} + G_{k'\uparrow}$. More clarified, lattice-pseudospin polarization may give rise to controlling the current of the junction to flow on A or B-sublattices. The SOI-region exhibits conductor with its carriers acquiring lattice-pseudospin $\Uparrow$ and $\Downarrow$.

**5. Filtering barrier using h-BN or SiC substrate (SSB-region)**

In this model, we have taken graphene grown on h-BN or SiC substrate as a lattice-pseudospin filtering barrier, because in this region it may behave like ***pseudo ferromagnet*** (PF) [21, 22] where the magnitude of pseudo-magnetization (PM) can be controlled by gate voltage. To calculate the expectation value of the lattice-pseudospin in SSB-region given by

$$\left\langle \vec{s}_{LP\text{-}spin} \right\rangle_{2,\eta s,\pm k} = \left\langle \psi_{2,\eta s,\pm k} \left| \hat{\vec{s}}_{LP\text{-}spin} \right| \psi_{2,\eta s,\pm k} \right\rangle.$$

where $\psi_{2,\eta s,\pm k} \approx \begin{pmatrix} 1 \\ \pm B e^{\pm i\eta\beta} \end{pmatrix} e^{\pm iqx + ik_\parallel y}$. We thus get the expectation value of lattice-pseudospin in the barrier of the from

$$\left\langle \vec{s}_{LP\text{-}spin} \right\rangle_{2,\eta s,\pm k} = \pm\eta\frac{\hbar}{2}\sqrt{1-\left(\frac{\Delta_{SSB}}{E-U_G}\right)^2}\left[\cos(\eta\theta)\hat{i} \pm \sin(\eta\theta)\hat{j}\right] + \frac{\hbar}{2}\frac{\Delta_{SSB}}{(E-U_G)}\hat{k}.$$

(16)

We will see that, in the limit of $|E - U_G|/\Delta_{SSB} \to 1$

$$\left\langle \vec{s}_{LP\text{-}spin} \right\rangle_{2,\eta s,\pm k} \to \text{sgn}(\Delta_{SSB}/[E-U_G])\frac{\hbar}{2}\hat{k} \propto PM_{\max}, \tag{17}$$

to get lattice-pseudospin only in one direction. This equation show that the barrier with SSB-gap would suppress lattice-pseudospin in some direction (see Figs.1(b) and 1(c)). The SSB-region plays the role of pseudo-ferromagnet acting on lattice-pseudospin direction. This property is applicable for lattice-pseudospin filtering device. We note that the formula above is done only for propagating state $|E - U_G| \geq \Delta_{SSB}$.

**6. Result and discussion**

In the numerical result, we investigate transport properties, by using the data given by ref.[39] for graphene grown on WS$_2$ substrate which may yield $\Delta_{SOI} = 5\,meV$, $\Delta_{SSB} = 0$ in SOI-region. Using the data given by ref.[27] for graphene grown on h-BN substrate which may yield $\Delta_{SOI} = 0$, $\Delta_{SSB} = 26.5\,meV$ in SSB-region. Moreover, in graphene grown on SiC substrate in ref.[29] may yield $\Delta_{SOI} = 0$, $\Delta_{SSB} = 250\,meV$ in the SSB-region.

We first study spin-valley conductances $G_{k\uparrow}, G_{k\downarrow}, G_{k'\uparrow}$ and $G_{k'\downarrow}$ as a function of gate voltage $U_G$ in Figs. 2(a) and (b). It is found that $G_{k\uparrow} = G_{k'\downarrow}$ and $G_{k\downarrow} = G_{k'\uparrow}$, leading to $G_\Uparrow \neq G_\Downarrow$. On the other hand, we find that $G_\uparrow = G_\downarrow$ and $G_k = G_{k'}$. It is said that the studied junction may give rise to the $L\text{-}PSP \neq 0$, while we get $VPS = 0$ and $SP = 0$. This behavior may be described by the role of lattice-pseudospin coupling to SOI-and SSB, discussed in sections 4 and 5. In the left SOI-region, electron of states $|k\uparrow\rangle, |k'\downarrow\rangle \equiv |\Uparrow\rangle$ and $|k\downarrow\rangle, |k'\uparrow\rangle \equiv |\Downarrow\rangle$ is injected into the pseudo-ferromagnetic barrier, SSB-region. The strongest pseudo-magnetization PM in the barrier may be proportional to expectation value of lattice-pseudospin in the z-direction for $|E - U_G| \geq \Delta_{SSB}$ at $|E - U_G| \to \Delta_{SSB}$, ie.,

$$PM \propto \left\langle \vec{s}_{z,LP\text{-}spin} \right\rangle = \frac{\hbar}{2} \frac{\Delta_{SSB}}{(E - U_G)} \hat{k}. \qquad (18)$$

The maximum value of $PM$ is found at $E - U_G = \pm\Delta_{SSB}$ where "$\pm$" denotes direction of PM in $\pm z$ axis. At this point $U_G = E \mp \Delta_{SSB}$, the junction would produce $L\text{-}PSP \approx \pm 100\%$ (see Fig3(a)). Linear control of $L\text{-}PSP$ as a function of gate potential for $|U_G| < \Delta_{SSB}$ from +100% to -100% is predicted. The slope does not change despite varying L. This behavior may be used to measure SSB-gap width in the junction. It is also found that only for $E/\Delta_{SOI} \to 1$, the first point that gives $L\text{-}PSP = 0$ is found at $U_G = \Delta_{SOI}$ (see Fig.3(a)). Also, this may be used to measure SOI in the junction, precisely. In Fig.3(b), we show that the lattice pseudospin polarization does not occur when increasing $E \gg \Delta_{SOI}$ because there is no transverse lattice-pseudo spin in SOI-regions (see eq.16). In this regime, pseudo magnetization goes to zero. The dips in $L\text{-}PSP$ found in Figs.3(a) and (b) may be described by tunneling resonance due to interference of Dirac electron wave inside the barrier when the wave vector $q = \sqrt{(E-U_G)^2 - (\Delta_{SSB})^2}/\hbar v_F$ is real. There is no dips of *L-PSP* found for $|E-U_G| < \Delta_{SSB}$ because $q$ is imaginary. As seen in Fig. 2(b) for L=100nm, only at the resonant points given for $|E-U_G| > \Delta_{SSB}$ the junction yields the same lattice-pseudospin conductance peaks. This leads to the dips of $L\text{-}PSP \to 0$ at the resonant points, given in Fig. 3. The period of the oscillation is proportional to $qL$. Thus, for a small thickness L=25nm the dips of *L-PSP* will be expected to occur when increasing $|U_G|$ greater than $50 meV$.

Finally, we study *L-PSP* as a function of gate potential at $L = 100\, nm$ and we take $E/\Delta_{SOI} \to 1$ in Fig.4, for various values of $\Delta_{SSB}$. As a result, the slope of linear control of *L-PSP* is found to adjustable by varying magnitude of SSB-gap in the barrier. Sharp switching of *L-PSP* from +100% to -100% is predicted for magnitude of $\Delta_{SSB}$ being comparable to the magnitude of $\Delta_{SOI}$. As seen in Fig.4, it may be concluded that the lattice-pseudospin polarization would be generated in the junction only for $\Delta_{SSB} \neq 0$. Since $\Delta_{SSB} = (E_A - E_B)/2$ where $E_{A(B)}$ is energy level in A(B) sublattice, it is possible to take $\Delta_{SSB}$ to be either positive or negative [24, 43, 44]. When $\Delta_{SSB}$ becomes negative, the *L-PSP* in Fig.4 will be switched to opposite sign, due to direction of PM being reversed (described by eq.18). The carriers with lattice-pseudospin opposite to the PM-direction would be suppressed. Changing sign of $\Delta_{SSB}$ thus just reverses the sign of *L-PSP*.

## 7. Summary and conclusion

We have investigated SOI/SSB/SOI gapped junction where SOI- and SSB-gapped graphene may be realized by growing graphene on WS$_2$ [39] and h-BN [27] or SiC-substrate [29], respectively. As a result, we found that the junction lead to control of lattice-pseudospin currents by gate when energy approaches spin-orbit interaction. The L-PSP changes linearly from 100% to -100% by varying gate potential and the slope of linear response may be adjusted by varying $\Delta_{SSB}$. Graphene grown on h-BN or SiC plays the role of pseudo-ferromagnet to cause lattice-pseudospin currents filtering while the presence of SOI produces the carriers as a source of lattice-pseudospin-dependent carriers. The SOI and SSB-gaps in graphene may be measured by characteristic of L-PSP as a function of gate potential. We also have discussed the mechanism to produced L-PSP in the junction by the presence of lattice-pseudospin coupling to SOI and SSB in graphene. Our proposed junction may be important for applications of graphene-based pseudospintroncs.

**Acknowledgment**

This research is supported in part by the Graduate Program Scholarship from the Graduate School, Kasetsart University.

**References**

[1] K.S. Novoselov, et al., Science **306** (2004) 666

[2] Y.-M. Lin et al., Science **327** (2010) 662

[3] F. Schwierz, Nature Nanotechnology **5** (2010) 487

[4] N. Kheirabadi, A. Shafiekhani, M. Fathipour, Superlattices and Microstructures **74** (2014) 123

[5] A. K. Geim, K. S. Novoselov, *Nature Materials* **6** (2007) 183

[6] P. R. Wallace, Phys. Rev. **71** (1947) 622

[7] H. Haugen et al., Physical Review B **77**(2008) 115406

[8] A. G. Swartz et al., ACS Nano **6** (2012) 10063

[9] E. Cobas et al., Nano Letters **12** (2012) 3000

[10] H. X. Yang et al., Physical Review Letters **110** (2013) 046603

[11] Z. Wang et al., Physical Review Letters **114** (2015) 016603

[12] T. Yokoyama, Phys. Rev. B **77**(2008) 073413

[13] J.-H. Park, H.-J. Lee, Physical Review B **89** (2014) 165417

[14] B. Soodchomshom et al., Physica E **41** (2009) 1310

[15] J. Bai et al., Nature Nanotechnology **5** (2010) 655

[16] Y. Song, G. Dai, Applied Physics Letters **106** (2015) 223104

[17] B. Soodchomshom, P. Chantngarm, Journal of Superconductivity and Novel Magnetism 24(2011)1885

[18] T. Fujita, M. B. A. Jalil, S. G. Tan, Applied Physics Letters **97**(2010) 043508

[19] M. M. Grujić et al., Physical Review Letters **113** (2014) 046601

[20] D. A. Pesin, A. H. MacDonald, Nature Materials **11** (2012) 409

[21] L. Majidi, M. Zareyan, Phys. Rev. B **83** (2011) 115422

[22] B. Soodchomshom, Physica E **44** (2012)1617

[23] L. Majidi, M. Zareyan, Journal of Computational Electronics **12** (2013) 134

[24] B. Soodchomshom et al., Physica E **52** (2013) 70

[25] P. San-Jose et al., Phys. Rev. Lett. 102 (2009) 247204

[26] S. Y. Zhou *et al., Nature Materials* **6** (2007) 916

[27] G. Giovannetti, et al., *Phys. Rev. B* **76** (2007) 073103

[28] R. Skomski et al., Mater. Horiz.1 (2014) 563

[29] M.S. Nevius et al., Phys. Rev. Lett. **115**(2015)136802

[30] D. Huertas-Hernando, et al., Phys. Rev. B **74** (2006)155426

[31] Y. Yao et al., Phys. Rev. B **75** (2007) 041401(R)

[32] H. Min, et al.,Phys. Rev. B **74** (2006)165310

[33] S. Konschuh, et al., Phys. Rev. B **82** (2010) 245412

[34] A. H. Castro Neto, F. Guinea, Phys. Rev. Lett. 103 (2009) 026804

[35] C. Weeks, et al., Phys. Rev. X 1 (2011)021001

[36] J. Balakrishnan, et al., Nat. Phys. 9 (2013) 284

[37] F. Calleja, et al., Nat. Phys. 11 (2015) 43

[38] A. Avsar, et al., Nat. Commun. **5**(2014)4875

[39] Z. Wang et al., Nat. Commun. **6**(2015)8339

[40] C. L. Kane, E. J. Mele Phys. Rev. Lett. **95** (2005) 226801

[41] M. M. Grujić et al., Phys. Rev. Lett. **113** (2014) 046601

[42] R. Landauer, IBM J. Res. Dev. 1 (1957) 223

[43] M. Zarenia, et al., Phys. Rev. B **86** (2012)085451

[44]  M.S. Fuhfer, Science **340** (2013) 1413

**Figure captions**

**Figure 1** Schematic illustration of (a) SOI/SSB/SOI-gapped graphene junction where SOI- and SSB-gapped graphene are provided by growing graphene on $WS_2$ and h-BN (SiC) substrate respectively. (b) The SSB-barrier exhibits pseudo-ferromagnet (PF) its magnetization (white arrows) being controlled by gate-voltage. To suppress some direction of lattice-pseudospin (dark arrows) in this junction leads to lattice-pseudospin filter.

**Figure 2** Plot of spin-valley conductances as a function of gate potential $U_G$ for (a) L=25 nm and (b) L=100nm.

**Figure 3** Plot of lattice-pseudospin polarisation (L-PSP) as a function of gate potential $U_G$ (a) for various values of L and (b) for various values of E.

**Figure 4** Plot of lattice-pseudospin polarisation (L-PSP) as a function of gate potential $U_G$ for various values of $\Delta_{SSB}$.

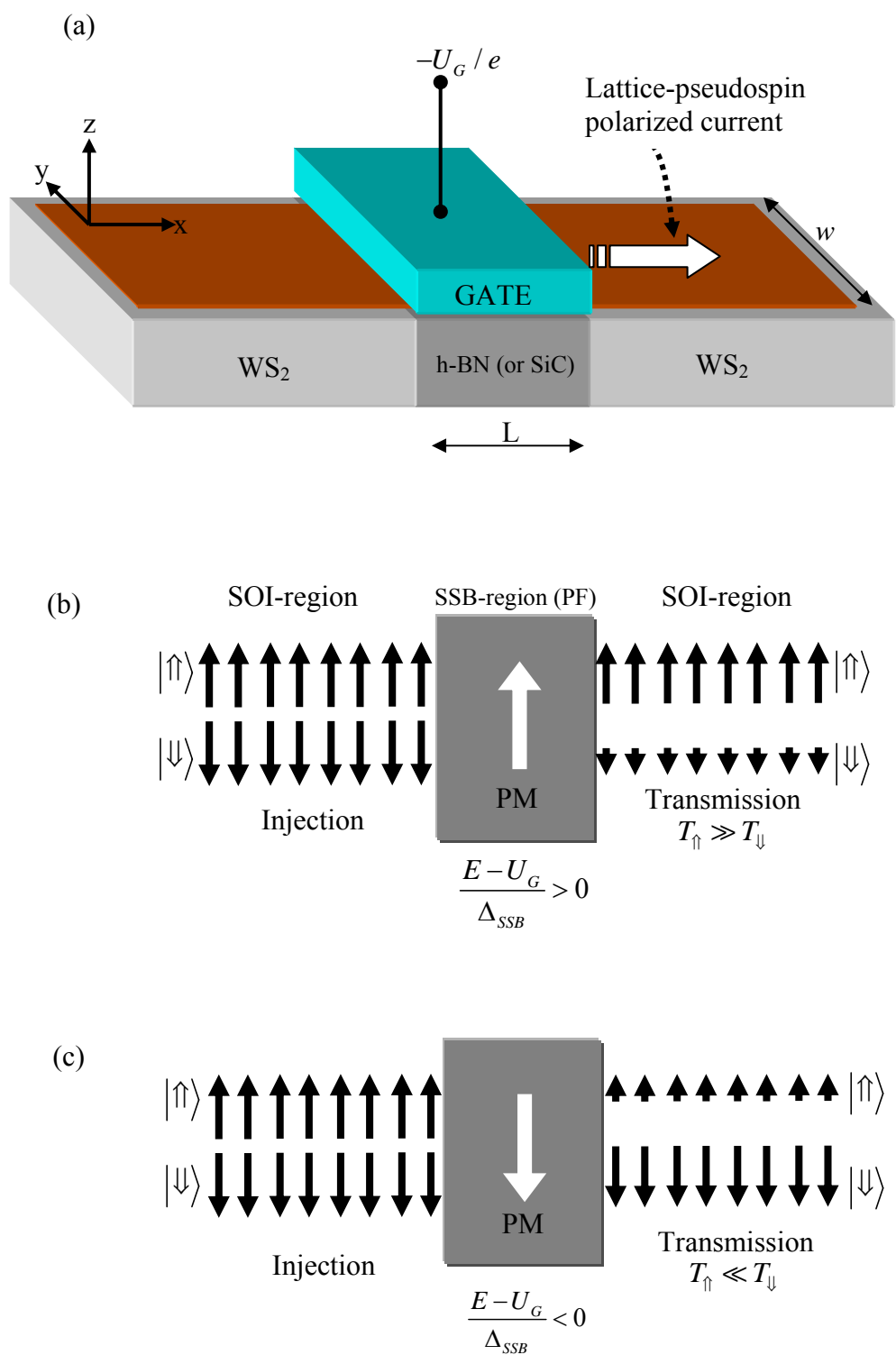

**Figure 1**

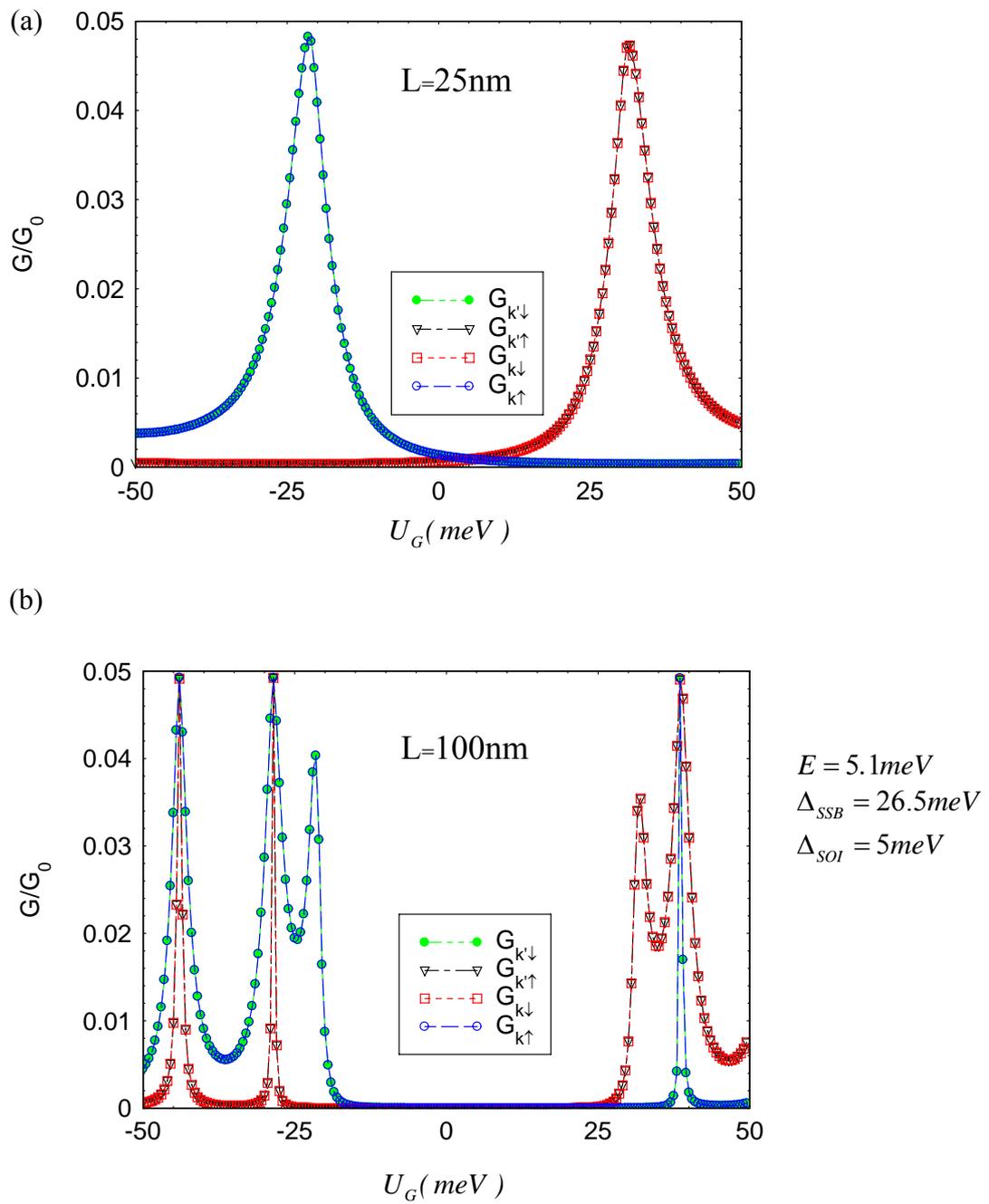

**Figure 2**

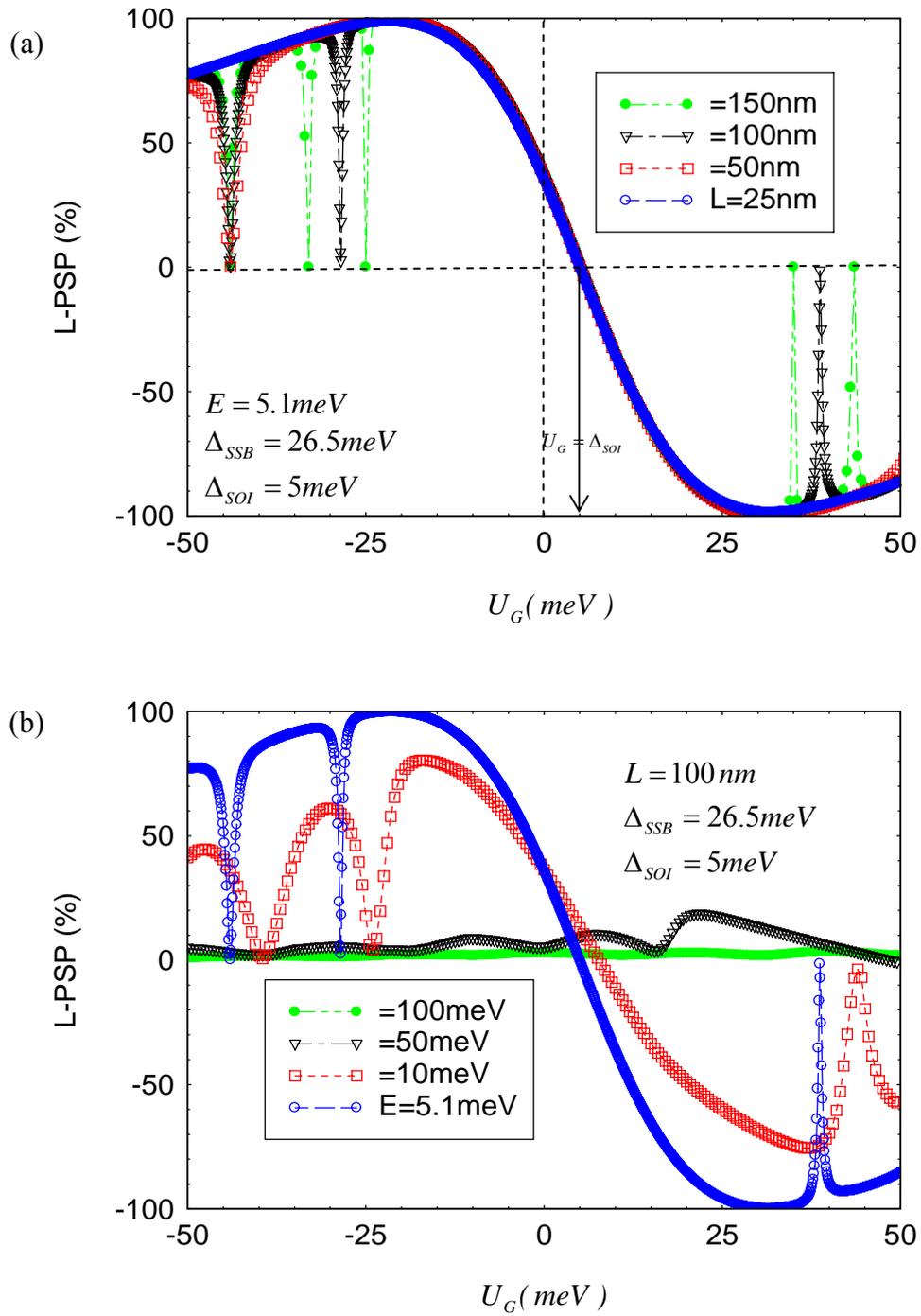

**Figure 3**

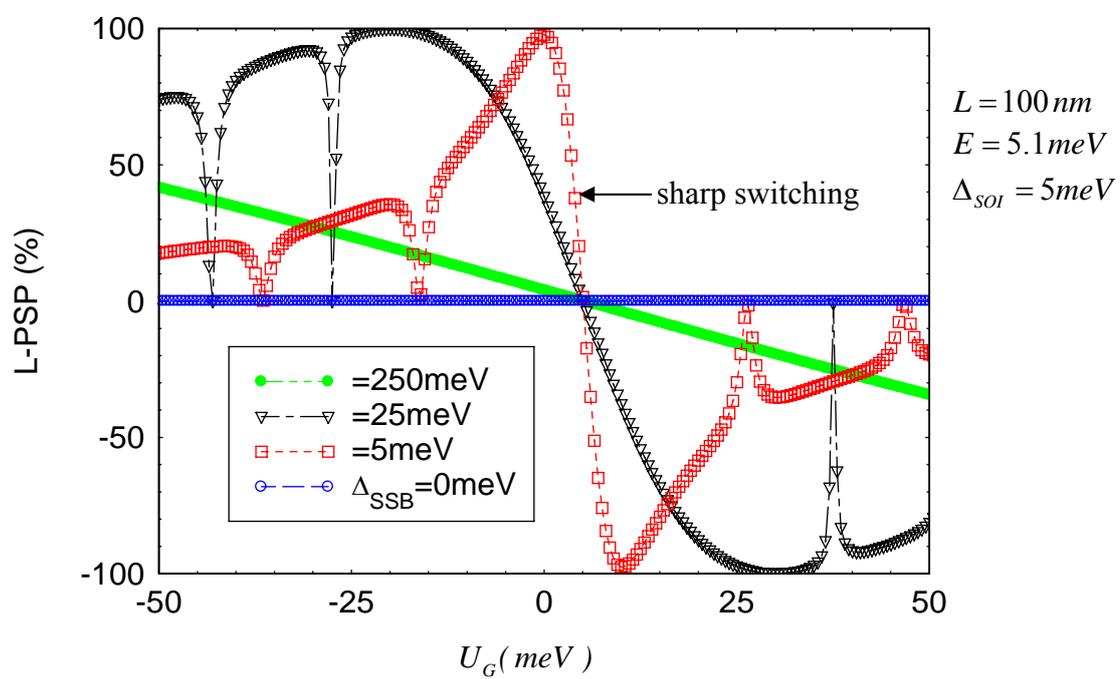

**Figure 4**